\documentclass[pre,unsortedaddress]{revtex4}%
\usepackage[english]{babel}
\usepackage[latin1]{inputenc}
\usepackage[dvips]{graphicx}
\usepackage{amsmath}
\usepackage{amssymb}
\usepackage{epsfig}
\usepackage{array}
\usepackage{amsfonts}%
\newcommand{\bel}[1]{\begin{equation}\label{#1}}

\def\bbbc{{\mathchoice {\setbox0=\hbox{$\displaystyle\rm C$}\hbox{\hbox
to0pt{\kern0.4\wd0\vrule height0.9\ht0\hss}\box0}}
{\setbox0=\hbox{$\textstyle\rm C$}\hbox{\hbox
to0pt{\kern0.4\wd0\vrule height0.9\ht0\hss}\box0}}
{\setbox0=\hbox{$\scriptstyle\rm C$}\hbox{\hbox
to0pt{\kern0.4\wd0\vrule height0.9\ht0\hss}\box0}}
{\setbox0=\hbox{$\scriptscriptstyle\rm C$}\hbox{\hbox
to0pt{\kern0.4\wd0\vrule height0.9\ht0\hss}\box0}}}}
\def\be{\begin{equation}}
\def\ee{\end{equation}}
\def\bege{\begin{equation}}
\def\ende{\end{equation}}
\def\bea{\begin{eqnarray}}
\def\eea{\end{eqnarray}}
\def\ba{\begin{array}}
\def\ea{\end{array}}

\begin{document}
\title{Infinite reflections of shock fronts in driven diffusive systems with two species}
\date{\today}
\author{V. Popkov }
\affiliation{Institut f\"ur Festk\"orperforschung, Forschungszentrum J\"ulich, 52425
J\"ulich, Germany }

\begin{abstract}
Interaction of a domain wall with boundaries of a system is studied for a
class of stochastic driven particle models. Reflection maps are introduced for
the description of this process. We show that, generically, a domain wall
reflects infinitely many times from the boundaries before a stationary state
can be reached. This is in an evident contrast with one-species models where
the stationary density is attained after just one reflection.

\end{abstract}
\maketitle






\section{Introduction}

\label{Intro}

There are many intrinsically nonequilibrium phenomena which can be observed
already in simplest systems of driven diffusing particles, a recent review of
which can be found e.g. in \cite{Schu00,Liggett1999,Gunter_reviewJPA}. Phase
transitions induced by spatial boundaries of a system is one of those
phenomena which was studied in detail for models with one-species of particles
\cite{Krug91,Kolo98,Gunter_Slava_Europhys} and for some multi-species models
\cite{Mukamel95,Peschel}. The ability of a nonequlibrium system to
\textquotedblleft feel\textquotedblright\ the boundaries constitutes a key
feature of driven systems: in fact, the boundaries dominate the bulk giving
rise to the phase transitions. It is a flux which brings information from the
boundaries to the bulk: in absence of a flux, the boundary conditions play
only a marginal role as one knows from equilibrium statistical mechanics.

A boundary problem in a one-dimensional driven diffusive system can be
formulated as follows: the system is coupled at the ends to reservoirs of
particles with fixed particle densities. A dynamics in the bulk and at the
boundaries is defined via hopping rates, which are time-independent. In this
case, after a certain transition period, the system will approach a stationary
state, characteristics of which (the average flux, the density profile, the
correlations) do not depend on time. It is of of interest how the system relax
to the stationary state. Since a stationary state is independent on an initial
state, one can choose an initial condition to be a domain wall interpolating
between the boundaries. It has long been recognized that boundary-driven phase
transitions are caused by the motion of a shock \cite{Kolo98}, so that the
above initial choice is a very natural one. A key issue is to understand how
the domain wall interacts with the boundaries. For the reference model in the
field, an Asymmetric Simple Exclusion Process (ASEP) with open boundaries, a
stationary density is reached after just one reflection, or interaction with a boundary.

We studied the reflection for the model with two conservation laws introduced
in \cite{Peschel,GunterJSP}, and suprisingly observed infinitely many
reflections occuring before a stationary state can be reached. More precisely,
a domain wall after the first right reflection (i.e., the interaction with the
right boundary) attains a different value of the density which is then changed
after the first left reflection, etc.. This process continues iteratively. The
bulk densities after multiple reflections asymptotically approach the
stationary values.

Description of iterative reflection maps constitutes the main subject of the
present paper.

Although the observed phenomenon of infinite reflections should be general, we
use for the study a class of driven diffusive models having a simple
stationary state on a ring for the sake of notational simplicity and to
eliminate possible source of errors which may result from approximating a bulk
flux, boundary conditions etc.. The findings are supported with Monte Carlo
calculations and hydrodynamic limit analysis.

The paper is organized as follows: In section~\ref{Model} we define the model
and its symmetries and describe the stationary state on a ring. Hydrodynamic
limit equations are given in section~\ref{PDE}. In section~\ref{Reflection}
reflection maps are introduced and infinite iterative sequences are analyzed.
Details concerning boundary rates can be found in the Appendix.

\section{The model}

\label{Model}

\begin{figure}[h]
\par
\begin{center}
\epsfig{width=7\unitlength,
file=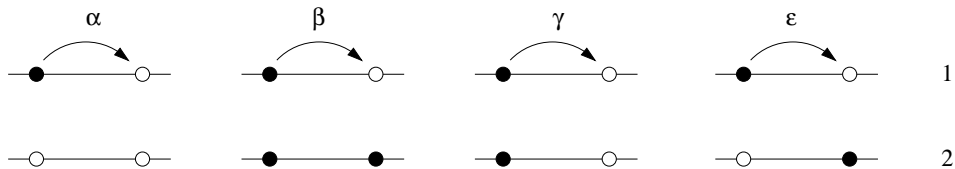}\vspace{3mm}
\end{center}
\caption{ The four elementary hopping processes, shown here for one of the
chains, and their rates. In the study, the rates are chosen to satisfy
(\ref{rates}).}%
\label{fig_4rates}%
\end{figure}

The model that we shall take as the example for our study is a particular case
of a more general model introduced in \cite{Peschel,GunterJSP}, which can be
viewed as a two-line generalization of ASEP. There are two parallel chains,
chain $A$ and chain $B$ consisting of $N$ sites each. Any site can be empty or
be occupied by a particle. The particle at a site $k$ can hop to its nearest
right-hand site $k+1$ on the same chain provided it is vacant. The rate of
hopping depends on the occupancy of the adjacent sites on the other chain, see
Fig.~\ref{fig_4rates}. Hopping between the chains is forbidden. The hopping
rates are fully symmetric with respect to an exchange between $A$ and $B$.
Correspondingly, there are 4 different rates $\alpha,\beta,\gamma,\epsilon$
(see Fig.~\ref{fig_4rates}). However if one chooses them to satisfy
$\alpha+\beta=2\gamma=2\epsilon$, then all particle configurations occur with
the same probability in a stationary state (see \cite{GunterJSP}). In this
case, the rate $\beta$ can be used to parametrize all the rates,
\begin{equation}
\alpha=1;\ \ \ \gamma=\epsilon=(1+\beta)/2. \label{rates}%
\end{equation}
Stationary fluxes on a ring with the densities $\rho^{A}$ and $\rho^{B}$ of
the particles on the $A$ and $B$ -chain respectively are given in the
thermodynamic limit (see \cite{GunterJSP}) by
\begin{equation}%
\begin{array}
[c]{cc}%
j^{A}(\rho^{A},\rho^{B}) & =\rho^{A}(1-\rho^{A})(1+(\beta-1)\rho^{B})\\
j^{B}(\rho^{A},\rho^{B}) & =\rho^{B}(1-\rho^{B})(1+(\beta-1)\rho^{A})
\end{array}
\label{fluxes}%
\end{equation}
For $\beta=1$ all the rates (\ref{rates}) become equal, which corresponds to
an absence of an interaction between the chains, and the system splits into
asymmetric exclusion processes \cite{ASEP,ASEP1,Liggett1999}. Another choice
of the rates $\alpha=\beta=\gamma\neq\epsilon$ was considered in
\cite{Peschel}.

The rates are by the definition nonnegative, so $\beta$ is in the range
$0\leq\beta<\infty$. However it is sufficient to consider $0\leq\beta\leq1$
because of a particle-hole symmetry. Indeed, an exchange of particles and
holes plus the substitution $\beta\rightarrow1/\beta$ leaves the model invariant.

In our context, we shall deal with a finite chains, where particles can be
injected at the left end and be extracted at the right end of the system. We
choose the boundary rates corresponding to effective stationary reservoirs of
particles with fixed particle densities. Explicit expressions for the boundary
rates are given, see the Appendix, in terms of the boundary densities. We
denote the densities of $A(B)$-particles in the left boundary reservoir as
$\rho_{L}^{A}(\rho_{L}^{B})$, and $\rho_{R}^{A}(\rho_{R}^{B})$ for the right
boundary reservoir.

\section{Hydrodynamic limit}

\label{PDE}

A naive continuum (Eulerian) limit of our stochastic dynamics on a lattice
${\hat{n}}_{k}(t)\rightarrow\rho^{\mathrm{A}}(x,t)$, ${\hat{m}}_{k}%
(t)\rightarrow\rho^{\mathrm{B}}(x,t)$ is a system of conservation laws
\begin{equation}
{\frac{\partial\rho^{Z}(x,t)}{\partial t}}+{\frac{\partial j^{Z}%
(\rho^{\mathrm{A}},\rho^{\mathrm{B}})}{\partial x}}=\varepsilon{\frac
{\partial^{2}\rho^{Z}}{\partial x^{2}}};\ \ \ Z=A,B, \label{cons2}%
\end{equation}

\[
\rho^{Z}(0,t)=\rho_{L}^{Z};\ \ \rho^{Z}(N,t)=\rho_{R}^{Z}%
\]
where on the right-hand side of (\ref{cons2}) a phenomenological vanishing
viscosity term has been added. This simplest regularization term leads to the
correct answer for the Riemann problem, as compared to the stochastic model,
see \cite{GunterJSP}, but fails to describe a reflection from the boundaries
(see Table~\ref{Table_derivative}). An adequate viscosity term is obtained by
averaging of exact lattice continuity equations of the stochastic process
\begin{align}
{\frac{\partial}{\partial t}}\hat{n}_{k}  &  =\hat{\jmath}_{k-1}^{\mathrm{A}%
}-\hat{\jmath}_{k}^{\mathrm{A}}\label{oper1}\\
{\frac{\partial}{\partial t}}\hat{m}_{k}  &  =\hat{\jmath}_{k-1}^{\mathrm{B}%
}-\hat{\jmath}_{k}^{\mathrm{B}} \label{oper2}%
\end{align}
for occupation number operators $\hat{n}_{k},\hat{m}_{k}$, and taking the
continuous limit $\langle\hat{n}_{k}\rangle\rightarrow\rho^{\mathrm{A}}(x,t)$,
$\langle\hat{m}_{k+1}\rangle\rightarrow\rho^{\mathrm{B}}(x+a,t)$. For the case
(\ref{rates}), the flux operator $\hat{\jmath}_{k}^{\mathrm{A}}$ reads (see
\cite{GunterJSP})%

\begin{equation}
\hat{\jmath}_{k}^{\mathrm{A}}=\hat{n}_{k}(1-\hat{n}_{k+1})\left(
1+{\frac{\beta-1}{2}}\left(  \hat{m}_{k}+\hat{m}_{k+1}\right)  \right)  ,
\label{flux_operator}%
\end{equation}
and $\hat{\jmath}_{k}^{\mathrm{B}}$ is obtained by an exchange $\hat
{n}\leftrightarrow\hat{m}$ in the above. We substitute (\ref{flux_operator})
into (\ref{oper1}),(\ref{oper2}), average, factorize and Taylor expand the
latter with respect to a lattice spacing $a\ll1$. In the first order of
expansion, one obtains%

\begin{align}
{\frac{\partial\rho^{\mathrm{A}}}{\partial t^{\prime}}}+{\frac{\partial
j^{\mathrm{A}}(\rho^{\mathrm{A}},\rho^{\mathrm{B}})}{\partial x}}  &
=\varepsilon{\frac{\partial}{\partial x}}\left(  \left(  1+(\beta
-1)\rho^{\mathrm{B}}\right)  {\frac{\partial\rho^{\mathrm{A}}}{\partial x}%
}\right) \label{hydroA}\\
{\frac{\partial\rho^{\mathrm{B}}}{\partial t^{\prime}}}+{\frac{\partial
j^{\mathrm{B}}(\rho^{\mathrm{A}},\rho^{\mathrm{B}})}{\partial x}}  &
=\varepsilon{\frac{\partial}{\partial x}}\left(  \left(  1+(\beta
-1)\rho^{\mathrm{A}}\right)  {\frac{\partial\rho^{\mathrm{B}}}{\partial x}%
}\right) \label{hydroB}\\
\varepsilon={\frac{a}{2}}\rightarrow0;\ \  &  {\frac{\partial}{\partial t}%
}=2\varepsilon{\frac{\partial}{\partial t^{\prime}}},
\end{align}
A numerical integration of (\ref{hydroA}),(\ref{hydroB}) shows that the
dynamics of the stochastic process is described adequately. At the same time,
predictions of (\ref{cons2}) strongly disagree with the stochastic dynamics
(see e.g. Table~\ref{Table_derivative}). Thus, in contradiction to the
one-species case, where a choice of the vanishing viscosity is very robust,
for a model with two and more species the specific choice of the viscosity
becomes crucial: different choices give different answers.

\section{Reflection maps}

\label{Reflection}

\begin{figure}[h]
\par
\begin{center}
\epsfig{width=7\unitlength,
file=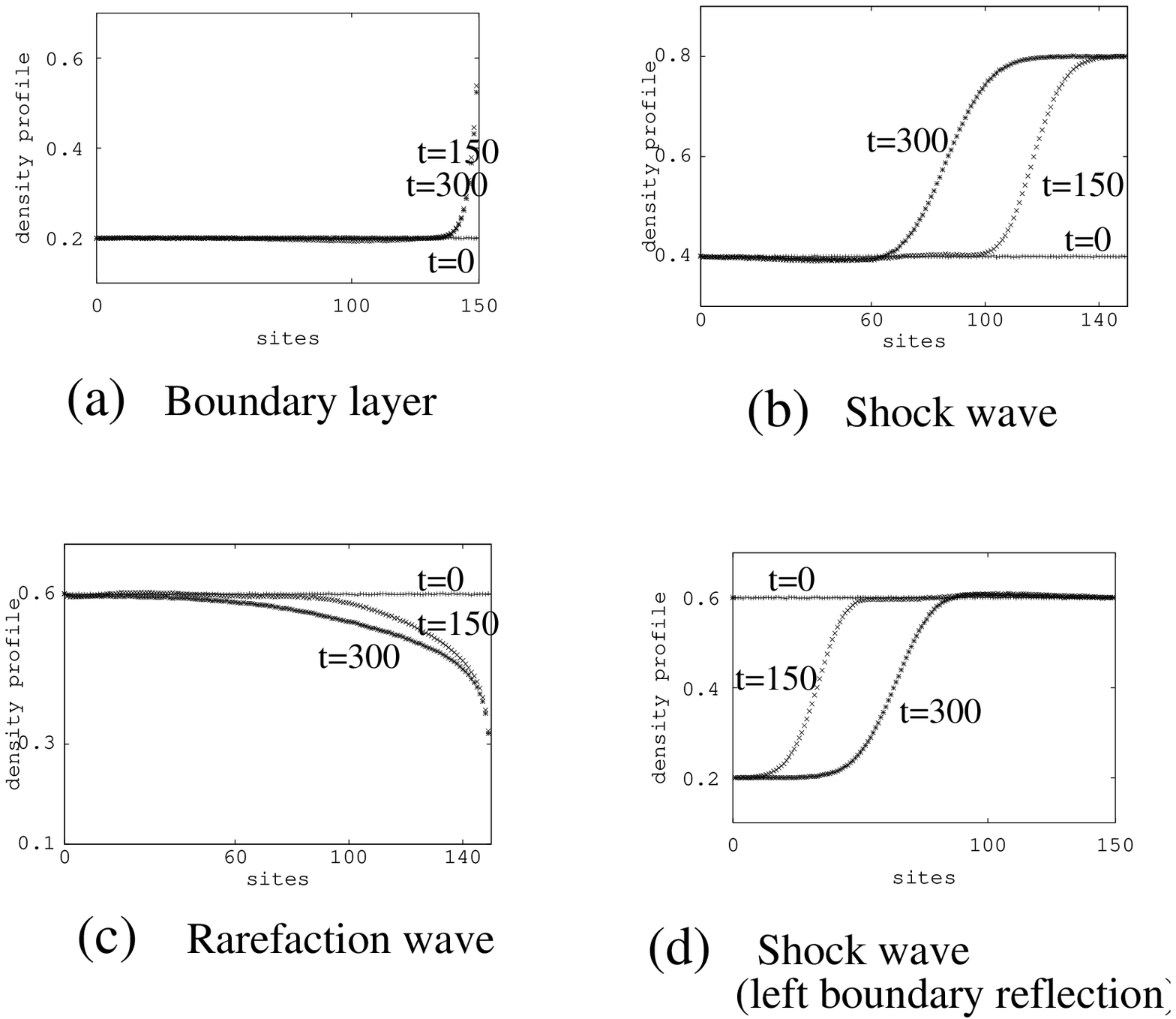}\vspace{3mm}
\end{center}
\caption{ Scenarios of a domain wall reflection from a boundary, as given by
Monte Carlo simulations. In all the cases, an initial distribution was
constant. The average profiles after $150$ and $300$ Monte Carlo steps are
shown. Averaging over $5\ast10^{5}$ histories is made. The graphs (a)-(c) show
the reflection from the right boundary. The graph (d) demonstrates the
reflection from the left boundary dual in the sense of (\ref{phs1}%
),(\ref{phs2}) to the graph (b)}%
\label{fig_scenariosASEP}%
\end{figure}

We explain our approach firstly for noninteracting chains $\beta=1$. In this
case the dynamics of each chain reduces to the one of the ASEP with an
injection rate $\rho_{L}$ and an extraction rate $(1-\rho_{R})$, see
\cite{ASEP,ASEP1}, corresponding to reservoirs of particles with densities
$\rho_{L}$ and $\rho_{R}$ on the left and on the right boundary, see the
Appendix. To study interaction with the right boundary, choose an initial
state to be a homogeneous state $\langle\hat{n}_{k}\rangle=\rho_{bulk}%
=\rho_{L}$, for all $k$, matching perfectly the left boundary. Consequently
the particle distribution at the left boundary will not change in time, while
at the right boundary there is a mismatch if $\rho_{L}\neq\rho_{R}$, which has
to be resolved. Monte-Carlo simulation shows that one of the following
scenarios is realized, see Fig.~\ref{fig_scenariosASEP}: (a) a thin boundary
layer develops interpolating between the bulk density $\rho_{\mathrm{bulk}%
}=\rho_{L}$ and $\rho_{R}$, and stays always attached to the right boundary,
(b) a shock wave of the density $\rho_{R}$ develops and propagates to the left
boundary, (c) a rarefaction wave with the density $\rho=max(\rho_{R},1/2)$
forms and spreads towards the left boundary.

Analogously, we study the left reflection, choosing an initial condition
$\rho_{L} \neq\rho_{bulk}=\rho_{R}$. Then, the mismatch will be only at the
left boundary. We can predict interaction results by exploiting a
particle-hole symmetry of the model: an exchange of particles and holes and
the left with the right boundary leaves the system invariant. Hence a dynamics
of the model with $\rho_{L} = \rho_{bulk} \neq\rho_{R}$ is equivalent to a
dynamics of the dual model
\begin{equation}
\label{phs1}\rho_{L}^{\prime}\neq\rho_{bulk}^{\prime}=\rho_{R}^{\prime},
\mathrm{where }\ \rho_{L}^{\prime}=1- \rho_{R}, \ \rho_{R}^{\prime}=1-
\rho_{L}:
\end{equation}
The corresponding local densities of the dual model satisfy
\begin{equation}
\label{phs2}\langle\hat n_{k} \rangle= \langle\hat{n^{\prime}}_{N-k+1}
\rangle; \ \ k=1,2, \ldots N.
\end{equation}
As an example, Fig.~\ref{fig_scenariosASEP}d shows the Monte Carlo evolution
corresponding to shock propagation, dual to the one on
Fig.~\ref{fig_scenariosASEP}b, through the transformations (\ref{phs1}%
),(\ref{phs2}). The other scenarios for the left reflection can be deduced
from Fig.~\ref{fig_scenariosASEP}a,c, using (\ref{phs1}),(\ref{phs2}).

We shall classify an outcome of a reflection from a boundary by an average
particle density $\rho_{refl}$ of the reflected wave measured at some point
$k^{*}$ not too close to the boundary to avoid boundary effects. The density
$\rho_{refl}$ is measured after the reflection wave has passed the point
$k^{*}$ but before it has reached the other boundary, to avoid possible
interference. For instance, reflected wave has the density $\rho_{refl}%
=\rho_{L}$ for the scenario (a), $\rho_{refl}=\rho_{R}$ for the scenario (b)
and $\rho_{refl}=1/2$ for the scenario (c) on Fig.~\ref{fig_scenariosASEP}.

We can summarize the results of all possible reflections from the right
boundary $\rho_{R}$ by plotting the density of the reflected wave versus the
initial bulk density, see Fig.~\ref{fig_reflASEP}. We shall call this type of
graph a reflection map. Reflection map for the left boundary is obtained using
(\ref{phs1}),(\ref{phs2}). Comparing the reflection maps with the stationary
phase diagram of ASEP \cite{ASEP,ASEP1}, one can make an important
observation. Namely, \textit{the stationary state density is achieved after
one reflection} from any of the boundaries. We have checked this statement to
be true also for a two-parameter model with a next nearest neighbour
interaction (KLS model), the stationary phase diagram of which is obtained in
\cite{Gunter_Slava_Europhys}.

In the following we show that for a two-component system it takes
\textit{infinite number of reflections} until the stationary density is reached.

\subsection{Two interacting chains}

Consider the two-chain model (\ref{rates}) coupled to boundary reservoirs with
the densities of $A$- and $B$- particles $\rho^{A}_{L},\rho^{B}_{L}$ and
$\rho^{A}_{R},\rho^{B}_{R}$ at the left and at the right end respectively.
Consider a reflection problem by choosing an initially homogeneous particle
distribution with the densities
\begin{equation}
\label{RR}\rho^{A}_{\mathrm{bulk}} = \rho^{A}_{L}; \ \ \rho^{B}_{\mathrm{bulk}%
} = \rho^{B}_{L} \ \ \mbox{ for the right reflection}
\end{equation}
\begin{equation}
\label{LR}\rho^{A}_{\mathrm{bulk}} = \rho^{A}_{R}; \ \ \rho^{B}_{\mathrm{bulk}%
} = \rho^{B}_{R} \ \ \mbox{ for the left reflection}
\end{equation}
We shall classify results of a reflection as it was done above for ASEP.
Denote the densities of the particles in the reflected wave in $A-$ and
$B$-chain as $r^{A},r^{B}$. Suppose that after interaction with a boundary an
interface appears between $\rho^{A}_{\mathrm{bulk}}, \rho^{B}_{\mathrm{bulk}}
$ and $r^{A},r^{B}$ moving towards the other boundary. Due to particle number
conservation in the bulk, a $Z$-component of the interface moves with the
velocity \cite{assume}
\begin{equation}
\label{v_shock}V^{Z} = {\frac{ j^{Z}_{\mathrm{bulk}} -j^{Z}_{\mathrm{r}}
}{\rho^{Z}_{\mathrm{bulk}}-r^{Z} }}\ ,
\end{equation}
where we used the short notations $j^{Z}_{\mathrm{bulk}} = j^{Z}(\rho
^{A}_{\mathrm{bulk}},\rho^{B}_{\mathrm{bulk}})$, $j^{Z}_{\mathrm{r}}
=j^{Z}(r^{A},r^{B})$. Since there is an interaction between the chains, the
velocities must coincide in both chains because a perturbation in one chain
causes the response in the other and vice versa. This gives a restriction
$V^{A} =V^{B}$, or
\begin{equation}
\label{vAvB}{\frac{ j^{A}_{\mathrm{bulk}} -j^{A}_{\mathrm{r}} }{\rho
^{A}_{\mathrm{bulk}}-r^{A} }} = {\frac{ j^{B}_{\mathrm{bulk}} -j^{B}%
_{\mathrm{r}} }{\rho^{B}_{\mathrm{bulk}}-r^{B} }},
\end{equation}
defining implicitly the allowed location of the points $r^{A}, r^{B}$.

For demonstration purposes, we choose $\beta=0$ in (\ref{rates}) which
corresponds to the maximal interchain interaction. This choice simplifies
drastically analytic expressions and at the same time preserves qualitative
features of the general case. For $\beta=0$, Eq.(\ref{vAvB}) has two families
of solutions, given by
\begin{align}
r^{A}/r^{B}  &  =\rho_{\mathrm{bulk}}^{A}/\rho_{\mathrm{bulk}}^{B}%
\label{solR}\\
(1-r^{A})(1-r^{B})  &  =(1-\rho_{\mathrm{bulk}}^{A})(1-\rho_{\mathrm{bulk}%
}^{B}) \label{solL}%
\end{align}
It turns out that the densities of waves resulting from the interaction with
the right boundary satisfy (\ref{solR}) while those resulting from the left
reflection satisfy (\ref{solL}).

\subsubsection{Right reflection}

\begin{figure}[h]
\par
\begin{center}
\epsfig{width=7\unitlength,
file=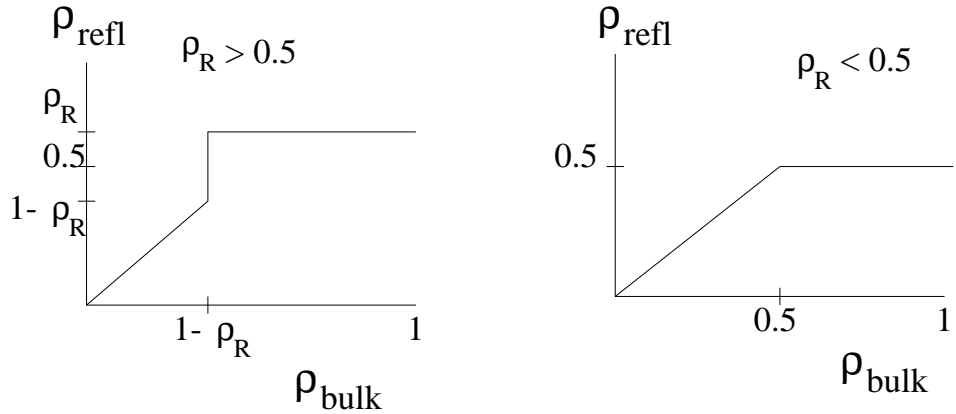}\vspace{3mm}
\end{center}
\caption{ The right boundary reflection map for ASEP }%
\label{fig_reflASEP}%
\end{figure}

\begin{figure}[h]
\par
\begin{center}
\epsfig{width=7\unitlength,
file=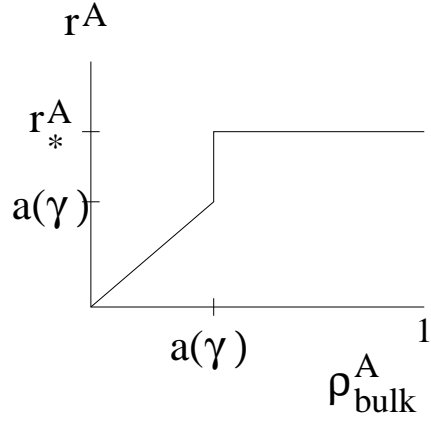}\vspace{3mm}
\end{center}
\caption{ Right boundary reflection map for the two-chain model along the line
${\rho^{B}_{\mathrm{bulk}}} / {\rho^{A}_{\mathrm{bulk}}}=\gamma$. The boundary
densities: $\rho^{A}_{R}=0.2, \rho^{B}_{R}=0.8$. Point $a(\gamma)$ denotes the
point of phase transition, where the bulk and the reflected fluxes become
equal $j^{Z}(a,\gamma a)=j^{Z}(r^{A}_{*},\gamma r^{A}_{*}) $. The locus of all
points $(r^{A}_{*},\gamma r^{A}_{*})$ and all points $a(\gamma),\gamma
a(\gamma)$ constitutes the upper and the bottom curve at
Fig.~\ref{fig_Refl_Right_All}, respectively. }%
\label{fig_refl_right_twocomponent}%
\end{figure}

Consider the right reflection (\ref{RR}) first. Monte Carlo simulations,
independently confirmed by a continuous model integration, lead to the
following conclusion:

the reflection densities $r^{A},r^{B}$ either coincide with the $\rho
^{A}_{\mathrm{bulk}} ,\rho^{B}_{\mathrm{bulk}}$ (in this case boundary layers
analogous to Fig.~\ref{fig_scenariosASEP}$a$ develop), or depend only on the
ratio $\gamma=\frac{\rho^{A}_{\mathrm{bulk}}} {\rho^{B}_{\mathrm{bulk}}}$. In
the latter case, $r^{B}(\gamma)/r^{A}(\gamma)=\gamma$, see (\ref{solR}). On
Fig.~\ref{fig_refl_right_twocomponent}, a typical dependence of $r^{A}$ on the
initial bulk density along the line $\frac{\rho^{B}_{\mathrm{bulk}}} {\rho
^{A}_{\mathrm{bulk}}}=\gamma$ is shown. Analogously to the left graph on
Fig.~\ref{fig_reflASEP}, a discontinuos change of $r^{A}$ occurs, defining a
point of a first order boundary driven phase transition \cite{other_case}.


Collection of points $r^{A}(\gamma), r^{B}(\gamma)$ for all possible initial
bulk densities $0\leq\gamma=\frac{\rho^{B}_{\mathrm{bulk}}} {\rho
^{A}_{\mathrm{bulk}}}\leq\infty$, constitutes a continuous curve $\mathcal{R}$
shown on Fig.~\ref{fig_Refl_Right_All}. The shape of $\mathcal{R}$ depends
only on the right boundary densities $\rho^{A}_{R},\rho^{B}_{R}$. In the
following we derive some analytical properties of $\mathcal{R}$, namely the
coordinates of three points on $\mathcal{R}$ and the derivative in one of them.

Trivially $\mathcal{R}$ contains the point $\rho_{R}^{A},\rho_{R}^{B}$
corresponding to the perfect match with the right boundary. Then, the location
of end points of $\mathcal{R}$ can also be derived. Indeed, consider the
system with the empty $B$-chain: $\rho_{\mathrm{bulk}}^{B}=0$. Then, the model
becomes effectively an ASEP, with the rate of extraction of the particles
$t_{\mathrm{ext}}^{0}$ given by (A.9), which corresponds to the effective
right boundary density in ASEP $\rho_{R}=1-t_{\mathrm{ext}}^{0}$. Consulting
the reflection graph for ASEP Fig.~\ref{fig_reflASEP}, we find that
corresponding reflected wave in the $A$-channel has the density $\max
(1-t_{\mathrm{ext}}^{0},0.5)$. Thus, the curve $\mathcal{R}$ has a point with
the coordinates $(\max(1-(1-\rho^{A}_{R})(1-\frac{1}{2}\rho_{R}^{B}%
),0.5),\ \ 0)$. Repeating the arguments for the empty $A$-chain $\rho
^{A}_{bulk}=0$, we find the other end point on $\mathcal{R}$, $(0,\ \ \max
(1-(1-\rho^{B}_{R})(1-\frac{1}{2}\rho_{R}^{A}),0.5))$.

Now, the exact derivative of $\mathcal{R}$ at the point $\rho_{R}^{A},\rho
_{R}^{B}$ can also be derived. Indeed, it is shown in the section~\ref{PDE}
that our system is described in the continuous limit by the system of
equations (\ref{hydroA},\ref{hydroB}). After the initial period of strong
interaction with the right boundary, the distribution of particles near the
boundary does not change anymore (see e.g. Fig.\ref{fig_scenariosASEP} b), and
can be described by a time-independent solution of (\ref{hydroA}%
,\ref{hydroB}). The current in the region will then be equal to that of the
reflected wave. Integrating the time-independent part of Eq.(\ref{hydroA}) for
$\beta=0$, one obtains:
\begin{equation}
j^{\mathrm{A}}(\rho^{\mathrm{A}}(x),\rho^{\mathrm{B}}(x))=j_{\mathrm{stat}%
}^{\mathrm{A}}(r^{\mathrm{A}},r^{\mathrm{B}})+\varepsilon{\frac{\partial
\rho^{\mathrm{A}}}{\partial x}}\left(  1-\rho^{\mathrm{B}}(x)\right)
\label{hydroA_stat}%
\end{equation}
where the current of the reflected wave $j_{\mathrm{stat}}^{\mathrm{A}%
}(r^{\mathrm{A}},r^{\mathrm{B}})$ is the constant of integration.

Suppose the result of the reflection $(r^{A},r^{B})$ differs only
infinitesimally from the right boundary values, $r^{Z}\approx\rho_{R}^{Z}$,
$\ Z=A,B$. The Taylor expansion of (\ref{hydroA_stat}) around the $r^{A}%
,r^{B}$ gives in the first order approximation
\begin{equation}
{\left(  {\frac{\partial j^{\mathrm{A}}}{\partial\rho^{A}}}\right)  }%
_{R}\delta\rho^{A}+{\left(  {\frac{\partial j^{\mathrm{A}}}{\partial\rho^{B}}%
}\right)  }_{R}\delta\rho^{B}=\lambda\left(  \delta\rho^{A}(1-\rho
_{R}^{\mathrm{B}})\right)  \label{mod_eig1}%
\end{equation}
where the subscript $R$ denotes evaluation at the point $\rho_{R}^{A},\rho
_{R}^{B}$, and we suppose the profile $\rho^{Z}(x)$ to approach exponentially
the values $r^{A},r^{B}$ from the right. An equation for the other flux
component $j^{\mathrm{B}}$, obtained analogously, can be combined together
with (\ref{mod_eig1}) into a modified eigen-value equation
\begin{equation}
{\left(  \mathcal{D}j\right)  }_{R}{\binom{\delta\rho^{A}}{\delta\rho^{B}}%
}=\lambda\mathcal{B}{\binom{\delta\rho^{A}}{\delta\rho^{B}}};\ \ \ \mathcal{B}%
=\left(
\begin{array}
[c]{cc}%
1-\rho_{R}^{\mathrm{B}} & 0\\
0 & 1-\rho_{R}^{\mathrm{A}}%
\end{array}
\right)  , \label{mod_eigen}%
\end{equation}
where $\mathcal{D}j$ is the Jacobian of the stationary flux (\ref{fluxes})
with the elements $\mathcal{D}_{UV}=\partial j^{U}/\partial\rho^{V}$. For any
given value of the boundary densities, $\lambda$ is determined
self-consistently, and the ratio $\delta\rho^{B}/\delta\rho^{A}$ is the
derivative of the curve $\mathcal{R}$ at the point $\rho_{R}^{A},\rho_{R}^{B}%
$. The system (\ref{mod_eigen}) has two solutions, one is relevant for the
right boundary reflection and one for the left reflection. The solution,
relevant for the right reflection, reads for $\beta=0$:
\begin{equation}
{\frac{\delta\rho^{B}}{\delta\rho^{A}}}={\frac{(1-\rho_{R}^{B})(\rho_{R}%
^{B}-\rho_{R}^{A}-\sqrt{(\rho_{R}^{A})^{2}+(\rho_{R}^{B})^{2}-\rho_{R}^{A}%
\rho_{R}^{B}})}{\rho_{R}^{A}(1-\rho_{R}^{A})}}. \label{sol}%
\end{equation}
We observe an excellent agreement between the theoretical and numerically
evaluated values of ${\frac{\delta\rho^{B}}{\delta\rho^{A}}}$, see
Table~\ref{Table_derivative}. The last column of the Table gives the value of
${\frac{\delta\rho^{B}}{\delta\rho^{A}}}$ which would follow from the
\textquotedblleft naive\textquotedblright\ viscosity term as written in
(\ref{cons2}), which leads correspondingly to $\mathcal{B}=I$ in
(\ref{mod_eigen}), and to the expression $({\frac{\delta\rho^{B}}{\delta
\rho^{A}})}_{naive}=-{\frac{1-\rho_{R}^{B}}{1-\rho_{R}^{A}}}$
\cite{Naive_case}. One sees from the Table that the latter choice gives a
wrong prediction.
\begin{table}[ptb]
\label{TableA}
\vspace{0.5cm}
\begin{tabular}
[c]{c|cccc}\hline
$( \rho^{A}_{R} \ \rho^{B}_{R} )$ & ${\frac{ \delta\rho^{B} }{\delta\rho^{A}}%
}$: & Theor & Numerical & ``Naive'' viscosity choice\\\hline
(0.2 0.8) &  & -0.151388 & -0.15(5) & -0.25\\
(0.3 0.8) &  & -0.190476 & -0.19(1) & -0.285714\\
(0.5 0.5) &  & -1 & -1 & -1\\
(0.4 0.6) &  & -0.548584 & -0.55(3) & -0.666667\\
(0. 0.6) &  & -0.2 & -0.2 & -0.4\\
(0.6 0.) &  & -5 & -5 & -2.5\\
($\rho^{A}$, 1) &  & 0 & 0 & 0\\\hline
\end{tabular}
\par
\caption{Comparison of the analytical formula (\ref{sol}) for derivative of
$\mathcal{R}$ at the point ($\rho_{R}^{A},\rho_{R}^{B}$) with the numerical
results.}%
\label{Table_derivative}%
\end{table}

\subsubsection{Left reflection}

\begin{figure}[h]
\par
\begin{center}
\epsfig{width=7\unitlength,
file=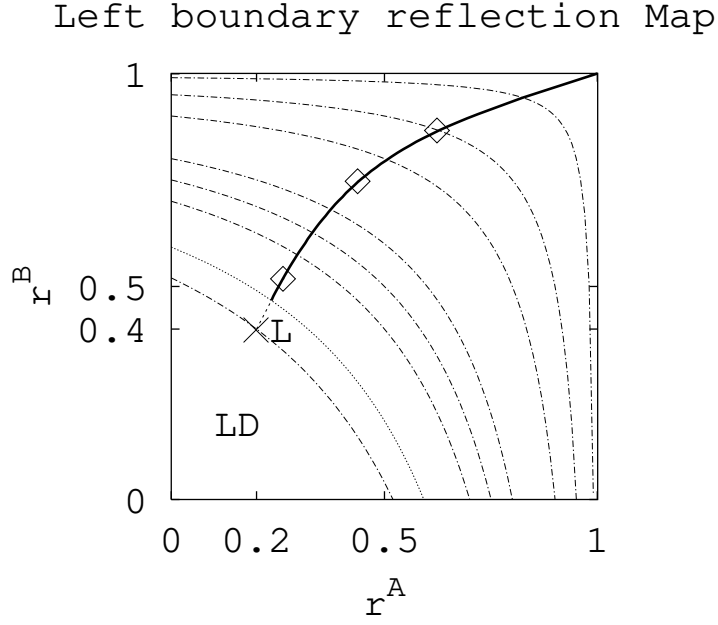}\vspace{3mm}
\end{center}
\caption{ Complete left boundary reflection map for the two-chain model. The
left boundary densities are: $\rho_{L}^{A}=0.2,\rho_{L}^{B}=0.4$. The bottom
left (LD) region marks the locus of initial densities resulting in the
reflection wave which matches the left boundary (point $L$ ). Squares show the
Monte-Carlo calculation points. Dotted curves show hyperbolas (\ref{gamma})
for different $\Gamma$. }%
\label{fig_Refl_Left_All}%
\end{figure}

Consider the reflection at the left boundary (\ref{LR}). We observe that the
densities of the reflected wave $r^{A},r^{B}$ either coincide with the left
boundary densities $\rho_{\mathrm{L}}^{A},\rho_{\mathrm{L}}^{B}$ or depend
only on the value $\Gamma$ characterising the respective solution
(\ref{solL})
\begin{equation}
(1-\rho_{\mathrm{bulk}}^{A})(1-\rho_{\mathrm{bulk}}^{B})=\Gamma\label{gamma}%
\end{equation}
More precisely, the curves (\ref{gamma}), $\ 0\leq\Gamma\leq1$ , see Fig.
\ref{fig_Refl_Left_All}, span the whole parameter space $0\leq\rho
_{\mathrm{bulk}}^{A},\rho_{\mathrm{bulk}}^{B}\leq1$ into three regions
$\Phi_{solid},\Phi_{dotted}$ (containing solid and dotted part of curve
$\mathcal{L}$ respectively), and $\Phi_{empty}$, containing the resting bottom
left part of ~Fig. \ref{fig_Refl_Left_All}. In the region $\Phi_{empty}$, the
wave matching perfectly left boundary reservoir, $\ \ (r^{A},r^{B}%
)=(\rho_{\mathrm{L}}^{A},\rho_{\mathrm{L}}^{B})$, is produced. In
$\Phi_{solid}$, reflection wave densities lie on $\mathcal{L}$ and for any
$\Gamma$ are given by an intersection of $\mathcal{L}$ with the hyperbola
(\ref{gamma}). Note that this intersection point is unique for any $\Gamma$.
Finally, in $\Phi_{dotted}$, the reflection wave consists of two consecutive
plateaus. First plateau from the right has the density $X^{A},X^{B}$ (point of
intersection between $\mathcal{L}$ with (\ref{gamma})), and the second matches
the left boundary. Since the velocity $V_{LX}$ of the interface between the
two is positive for all $X^{A},X^{B}$ belonging to the dotted part of
reflection curve $\mathcal{L}$,
\begin{equation}
V_{LX}=\frac{j^{A}(X^{A},X^{B})-j^{A}(\rho_{\mathrm{L}}^{A},\rho_{\mathrm{L}%
}^{B})}{X^{A}-\rho_{\mathrm{L}}^{A}}=\frac{j^{B}(X^{A},X^{B})-j^{B}%
(\rho_{\mathrm{L}}^{A},\rho_{\mathrm{L}}^{B})}{X^{B}-\rho_{\mathrm{L}}^{B}}>0,
\label{VLX}%
\end{equation}
the effective reflection result is $(r^{A},r^{B})=(\rho_{\mathrm{L}}^{A}%
,\rho_{\mathrm{L}}^{B})$ like in the region $\Phi_{empty}$. At the point $K$
where the dotted and solid part of the curve $\mathcal{L}$ meet, the velocity
(\ref{VLX}) vanishes, $V_{LX}(K)=0$. Inside the region $\Phi_{solid}$ the
velocity $V_{LX}<0$ meaning that the intermediate shock interface stays
\textquotedblleft glued\textquotedblright\ to the left boundary and only one
reflection wave survives. Note also that due to (\ref{VLX}), the points
$X^{A},X^{B}$ from the dotted part of $\mathcal{L}$ satisfy $X^{A}/X^{B}%
=\rho_{\mathrm{L}}^{A}/\rho_{\mathrm{L}}^{B}$ as follows from (\ref{VLX}%
),(\ref{solR}).

Generically, the curve $\mathcal{L}$, corresponding to arbitrary boundary
densities $\rho_{\mathrm{L}}^{A},\rho_{\mathrm{L}}^{B}$ , contains this point
itself, corresponding to a perfect match with the left boundary. Secondly, it
ends at the point $(1,1)$ for the case $\beta=0$ which we consider. Indeed, if
$\rho_{\mathrm{bulk}}^{B}\approx1$, the chain $B$ is almost completely full,
so that no flux can flow at the chain $B$ because of exclusion rule. But, a
flux is very small also in the chain $A$, because the dominating
configurations, see Fig.\ref{fig_4rates}, have zero bulk hopping rate
$\beta=0$. On the other hand, the particles can enter with a finite rate
(\ref{inj_1A}). In the limit $\rho_{\mathrm{bulk}}^{B}\rightarrow1$ this
yields the reflected densities $(r^{A},r^{B})=(1,1)$. Or, in terms of the
reflection map, $(\rho_{\mathrm{bulk}}^{A},1)\rightarrow(1,1)$ Exchanging the
role of the $A$- and $B$-chain leads to the same result, $(1,\rho
_{\mathrm{bulk}}^{B})\rightarrow(1,1)$, due to complete symmetricity of the rates.

\begin{figure}[pbh]
\caption{ Complete right boundary reflection map for the two-chain model,
showing the locus of the allowed densities of the waves reflected from the
right boundary. The right boundary densities are: $\rho^{A}_{R}=0.2, \rho
^{B}_{R}=0.8$. Squares show the Monte-Carlo calculation points, the upper
continuous curve $\mathcal{R}$ is obtained by integrating the hydrodynamic
equation (\ref{hydroA},\ref{hydroB}). The crossmark shows the right boundary
densities. The bottom left (LD) region marks the low density phase and the
bottom curve marks the line of the first order phase transition, see also
Fig.~\ref{fig_refl_right_twocomponent}. }%
\label{fig_Refl_Right_All}
\par
\begin{center}
\epsfig{width=7\unitlength,
file=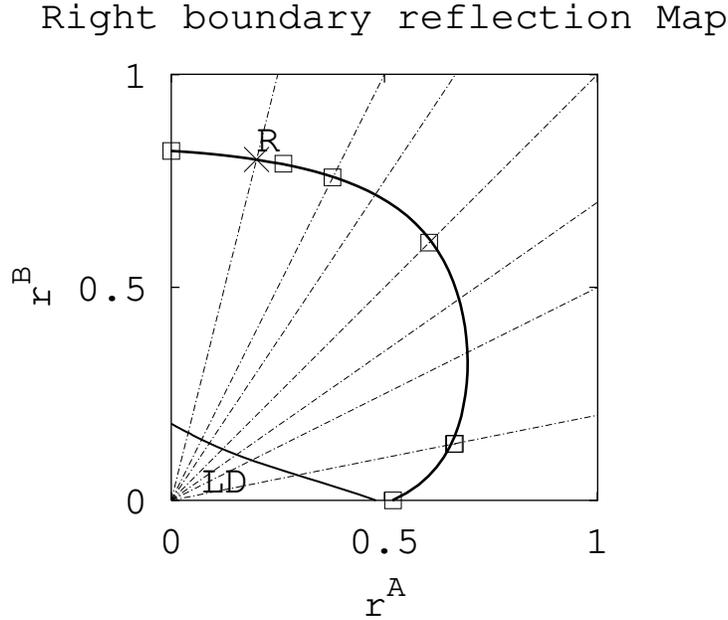}\vspace{3mm}
\end{center}
\end{figure}

The reflection maps Figs.\ref{fig_Refl_Right_All},\ref{fig_Refl_Left_All} can
be interpreted also in the following sense: the right reflection map
Fig.~\ref{fig_Refl_Right_All} shows the location of all possible stationary
densities of a system with the given right boundary densities $\rho
_{\mathrm{R}}^{A},\rho_{\mathrm{R}}^{B}$. Indeed, after a right reflection the
density of particles in the system is located either on $\mathcal{R}$ or in
the bottom left (low density) region. Correspondingly, a stationary state
(whatever are the conditions on the left boundary) must belong to the same
domain. Analogously, the left reflection map Fig.~\ref{fig_Refl_Left_All},
considered alone, shows that the stationary densities of the system with the
given left boundary densities either lie on the curve $\mathcal{L}$ or match
the left boundary $\rho_{\mathrm{L}}^{A},\rho_{\mathrm{L}}^{B}$.

\begin{figure}[h]
\par
\begin{center}
\epsfig{width=7\unitlength,
file=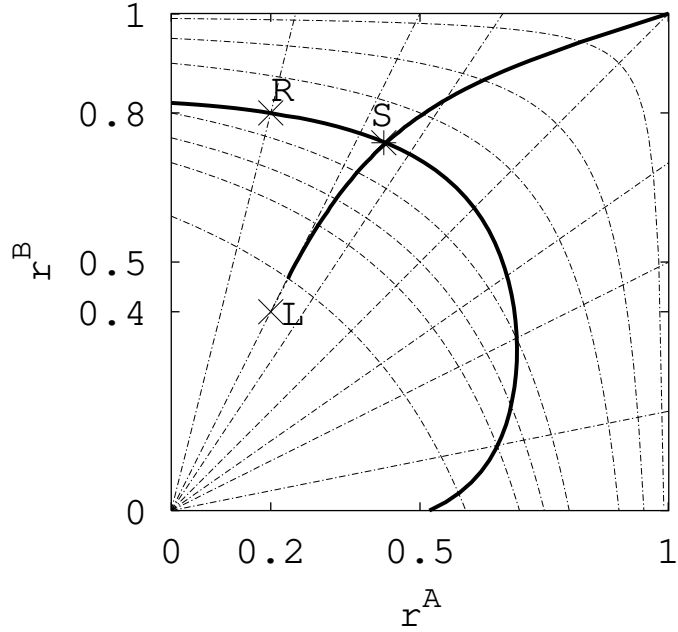}\vspace{3mm}
\end{center}
\caption{ Curves $\mathcal{R}$ and $\mathcal{L}$ from the
Figs.~\ref{fig_Refl_Right_All},\ref{fig_Refl_Left_All} combined together.
Their crossing, $\mathcal{S}$, is the stationary state density achieved
through the infinite number of reflections. The crossmarks near the symbols
$\mathcal{R}$ and $\mathcal{L}$ indicate the left and right boundary
densities, and the star at the crossing --- the stationary density obtained by
Monte Carlo simulation of the system of $300$ sites. The systems was
equilibrated for = $4*10^{5}$ Monte Carlo Steps (MCS), after which the
averaging over MCS = $4*10^{5}$ and $10$ different histories was done. }%
\label{fig_2RL}%
\end{figure}

\begin{figure}[h]
\par
\begin{center}
\epsfig{width=5\unitlength,
file=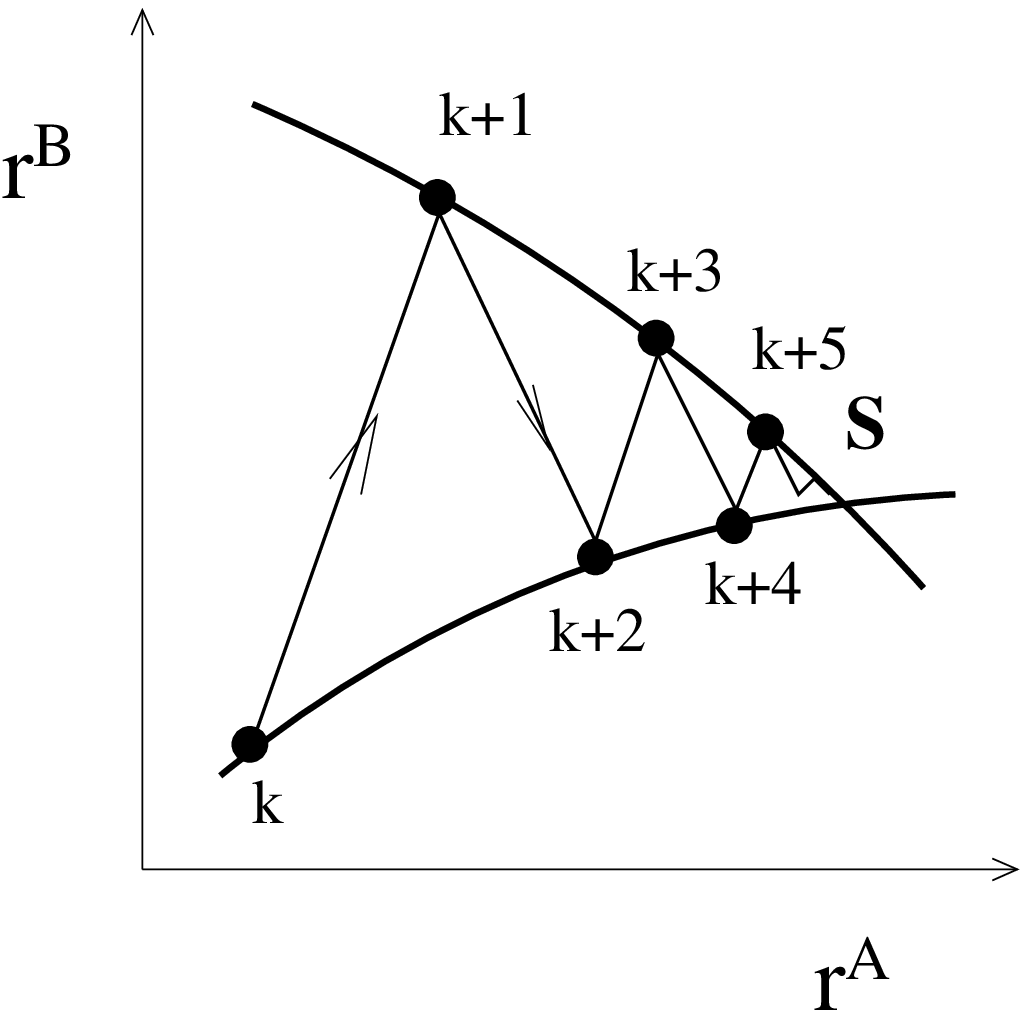}\vspace{3mm}
\end{center}
\caption{ Approach to stationary point $\mathcal{S}$ through an infinite
sequence of reflections. The numerated filled circles show the location of
subsequent domain wall densities $r^{A},r^{B}$. The circles $k,k+2,k+4$ lie on
the curve $\mathcal{L}$ and correspond to the result of the left reflection,
while those lying on the upper curve $\mathcal{R}$, correspond to the right
reflection}%
\label{fig_enlarge_2RL}%
\end{figure}

We shall not explore further the analytical properties of the curves
$\mathcal{L}$ and $\mathcal{R}$, but investigate what happens if we fix the
right and the left boundary corresponding to the reflection maps
Fig.~\ref{fig_Refl_Right_All} and Fig.~\ref{fig_Refl_Left_All}. Initially
empty system fills with the particles from the left end, with the densities,
fitting the density of the left reservoir, Fig.~\ref{fig_Refl_Left_All}. Thus
the wave $\rho_{\mathrm{bulk}}^{A},\rho_{\mathrm{bulk}}^{B}=\rho_{L}^{A}%
,\rho_{L}^{B}$ starts to propagate in the system towards the right boundary.
Reaching the boundary, a reflected wave forms, with the densities $r_{0}^{A}$
,$r_{0}^{B}$ \ given by the intersection of the line $y/x=\rho_{L}^{B}%
/\rho_{L}^{A}$ with the curve $\mathcal{R}$ of Fig.~\ref{fig_Refl_Right_All}.
This reflected wave hits the left boundary then, resulting in the next
reflected wave with the densities $r_{1}^{A}$ ,$r_{1}^{B}$. The left
reflection is controlled by the curve $\mathcal{L}$ at
Fig.~\ref{fig_Refl_Left_All}, therefore the densities of the reflected wave
$r_{1}^{A}$ ,$r_{1}^{B}$ are given by the intersection of hyperbola
$(1-y)(1-x)=(1-r_{0}^{A})(1-r_{0}^{B})$ with the curve $\mathcal{L}$. Since
the curves $\mathcal{R}$,$\mathcal{L}$ do not coincide with any of the curves
defined by (\ref{solL},\ref{solR}) in our example, this process continues
forever, though converging to stationary state $\mathcal{S}$, the point of the
intersection of $\mathcal{L}$ with $\mathcal{R}$, see Fig.\ref{fig_2RL}. After
$2k-1$ reflections, the wave with the densities $r_{2k-1}^{A}$ ,$r_{2k-1}^{B}$
will hit the right boundary, producing the reflected wave of the densities
$r_{2k}^{A}$ ,$r_{2k}^{B}$, corresponding to intersection of the line
$y/x=r_{2k-1}^{A}/r_{2k-1}^{B}$ with the curve $\mathcal{R}$. The latter wave,
hitting the left boundary, produces the next reflected wave with the densities
$r_{2k+1}^{A}$ ,$r_{2k+1}^{B}$ , the point of the intersection of hyperbola
$(1-y)(1-x)=(1-r_{2k}^{A})(1-r_{2k}^{B})$ with $\mathcal{L}$. This process is
depicted schematically on Fig.\ref{fig_enlarge_2RL}. Several remarks are in order.

\begin{itemize}
\item The densities $\ \{r_{k}^{A}$ ,$r_{k}^{B}\}_{k=0}^{\infty}$ of the
reflected waves constitute converging sequence. Convergence is exponential,
$\delta u_{n+2k}=e^{-\kappa k}\delta u_{n}$, for $n\rightarrow\infty$ where we
denote by $\delta u_{n}$ the deviation from the stationary density at the
$n$-th step. $e^{-\kappa}<1$ is a constant depending on tangential derivatives
$\alpha,\beta,\alpha_{1},\beta_{1}$ of the curves $\mathcal{R}$,$\mathcal{L}$
and the characteristic curves, respectively, at the stationary point
$\mathcal{S}$: $\ e^{-\kappa}=(1-\frac{tg(\alpha_{1})}{tg(\alpha)}%
)(1-\frac{tg(\beta_{1})}{tg(\beta)})/((1+\frac{tg(\alpha_{1})}{tg(\beta
)})(1-\frac{tg(\beta_{1})}{tg(\alpha)}))$. Note that if \ at least one of the
characteristic derivatives happens to coincide with $\alpha_{1},\beta
_{1}:\alpha=\alpha_{1}$ or $\beta=\beta_{1}$ then the sequence converges in
one step.

\item The velocities of the reflected wave for big $k$ converge to the finite
characteristic velocities computed at the stationary point $\mathcal{S}$, as
eigenvalues of the flux Jacobian, see \cite{GunterJSP}.

\item The stationary densities are not reached at any finite time
\cite{strictly_speaking}.
\end{itemize}

We would like to stress that convergence to a stationary state through
infinite number of reflections is a generic feature of a two-component model.
However, in specific cases the convergence to stationary state can be achieved
after finite number of reflections. E.g., if the left boundary densities are
equal, the curve $\mathcal{L}$ will be a straight line $y=x$ coinciding with
the one of the solutions (\ref{solR}), and consequently the corresponding
stationary state will be reached in one step, like in the one-species case.

\section{Conclusion}

To conclude, we have studied two lane particle exclusion processes on the
microscopic level, focusing on interaction of domain walls with the boundaries
of the system. It was shown that a shock front interacts infinite number of
times with the boundaries of the system in order to establish a stationary
state density. It was also shown that hydrodynamic limit with a usually taken
approximation of diagonal vanishing viscosity ( see, e.g., \cite{Bressan})
fails to describe the microscopic dynamics. An adequate hydrodynamic limit is constructed.

For the description of the reflection of domain walls from the boundaries, we
introduced reflection maps, which show the locus of the allowed densities of
the reflected waves. Careful analysis of such maps is necessary in order to
solve a boundary problem for models with more than one species of particles.

\begin{acknowledgments}
I thank A.Rakos, G.M. Sch\"{u}tz and M. Salerno for fruitful discussions, and
acknowledge financial support from the Deutshe Forschungsgemeinschaft and the
hospitality of the Department of Theoretical Physics of the University of
Salerno where a part of the work was done.
\end{acknowledgments}

\appendix* \label{A}

\section{Boundary rates}

Here we consider our model on two parallel chains of length $N$ where constant
densities of particles are kept on the left and on the right boundary. This
can be achieved by choosing appropriately boundary rates for injection and
extraction of the particles.

The average current through the left boundary of chain $A$ is by definition
\begin{equation}
j_{L}^{A}=r_{\mbox{inj}}^{1}(A)\langle(1-\hat{n}_{1})\hat{m}_{1}%
\rangle+r_{\mbox{inj}}^{0}(A)\langle(1-\hat{n}_{1})(1-\hat{m}_{1})\rangle,
\label{jAL}%
\end{equation}
where $r_{\mbox{inj}}^{1(0)}(A)$ is the probability to inject a particle on
the first site $k=1$ of chain $A$ provided adjacent site is occupied (empty).
On the other hand, the average current through the link between the sites $k$
and $k+1$ of chain $A$ is given by
\begin{align}
j^{A}  &  =1\langle\hat{n}_{k}(1-\hat{m}_{k})(1-\hat{n}_{k+1})(1-\hat{m}%
_{k+1})\rangle+\beta\langle\hat{n}_{k}\hat{m}_{k}(1-\hat{n}_{k+1})\hat
{m}_{k+1}\rangle\nonumber\label{jA}\\
&  +{\frac{\beta+1}{2}}\langle\hat{n}_{k}\hat{m}_{k}(1-\hat{n}_{k+1}%
)(1-\hat{m}_{k+1})\rangle+{\frac{\beta+1}{2}}\langle\hat{n}_{k}(1-\hat{m}%
_{k})(1-\hat{n}_{k+1})\hat{m}_{k+1}\rangle
\end{align}
Let the left boundary density to be $\rho_{L}^{A}(\rho_{L}^{B})$ for chains
$A$ $(B)$ respectively and assume the same stationary density in the bulk.
Then the bulk current and the boundary current must be equal $j_{L}^{A}=j^{A}%
$. Comparing (\ref{jAL}) and (\ref{jA}), and accounting for the absence of
correlations in the stationary state, see \cite{GunterJSP}, we obtain the
injection rates
\begin{equation}
r_{\mbox{inj}}^{1}(A)=\beta\langle\hat{n}\rangle\langle\hat{m}\rangle
+{\frac{\beta+1}{2}}\langle\hat{n}\rangle\langle1-\hat{m}\rangle=\rho_{L}%
^{A}\left(  1+{\frac{\beta-1}{2}}(1+\rho_{L}^{B})\right)  \label{inj_1A}%
\end{equation}

\begin{equation}
\label{inj_0A}r_{\mbox{inj}}^{0} (A) = \langle\hat n \rangle\langle1-\hat m
\rangle+ {\frac{\beta+1 }{2}} \langle\hat n \rangle\langle\hat m \rangle=
\rho^{A}_{L} \left(  1+ {\frac{\beta-1 }{2}} \rho^{B}_{L} ) \right)
\end{equation}

Now, due to the symmetricity of the model, the injection rates $r_{\mbox{inj}}%
^{1}(B)$,$r_{\mbox{inj}}^{0}(B)$ for the chain $B$ are obtained by an exchange
of $\rho_{L}^{A}\leftrightarrow\rho_{L}^{B}$ in the above expressions:
\begin{equation}
r_{\mbox{inj}}^{1}(B)=\rho_{L}^{B}\left(  1+{\frac{\beta-1}{2}}(1+\rho_{L}%
^{A})\right)
\end{equation}%
\begin{equation}
r_{\mbox{inj}}^{0}(B)=\rho_{L}^{B}\left(  1+{\frac{\beta-1}{2}}\rho_{L}%
^{A})\right)
\end{equation}
Analogously, the current through the right boundary is due to the extraction
of the particles from the right end $k=N$ of the chains $A$ and $B$,%

\begin{equation}
\label{jAR}j^{A}_{R} = t_{\mbox{ext}}^{0} (A) \langle\hat n_{N} (1-\hat
m_{N})\rangle+ t_{\mbox{ext}}^{1} (A) \langle\hat n_{N} \hat m_{N} \rangle,
\end{equation}
\begin{equation}
\label{jBR}j^{B}_{R} = t_{\mbox{ext}}^{0} (B) \langle\hat m_{N} (1-\hat
n_{N})\rangle+ t_{\mbox{ext}}^{1} (B) \langle\hat n_{N} \hat m_{N} \rangle.
\end{equation}
Assuming the right boundary densities (and the bulk densities) to be $\rho
^{A}_{R},\rho^{B}_{R}$ and equating (\ref{jAR}) with (\ref{jA}), we obtain%

\begin{equation}
t_{\mbox{ext}}^{0}(A)=\beta\langle1-\hat{n}\rangle\langle1-\hat{m}%
\rangle+{\frac{\beta+1}{2}}\langle1-\hat{n}\rangle\langle\hat{m}%
\rangle=(1-\rho_{R}^{A})\left(  1+{\frac{\beta-1}{2}}\rho_{R}^{B}\right)
\end{equation}%
\begin{equation}
t_{\mbox{ext}}^{1}(A)=(1-\rho_{R}^{A})\left(  1+{\frac{\beta-1}{2}}(1+\rho
_{R}^{B})\right)  .
\end{equation}
The expressions for the extraction rates from the chain $B$ are obtained by
exchanging $\rho_{L}^{A}\leftrightarrow\rho_{L}^{B}$ in the above,
\begin{equation}
t_{\mbox{ext}}^{0}(B)=(1-\rho_{R}^{B})\left(  1+{\frac{\beta-1}{2}}\rho
_{R}^{A}\right)
\end{equation}%
\begin{equation}
t_{\mbox{ext}}^{1}(B)=(1-\rho_{R}^{B})\left(  1+{\frac{\beta-1}{2}}(1+\rho
_{R}^{A})\right)  .
\end{equation}

For any fixed choice of boundary densities the model will relax to unique
stationary state, independent on the initial conditions. For equal left and
right boundary densities, $\rho^{A}_{R} = \rho^{A}_{L}=\rho^{A}$, $\rho
^{B}_{R} = \rho^{B}_{L}= \rho^{B}$, our definition of the boundary rates
guarantees that the stationary state will be the uncorrelated homogenuous (
product) state $\langle\hat n_{k} \rangle= \rho^{A}$, $\langle\hat m_{k}
\rangle= \rho^{B}$ for \textit{any} number of sites $N>1$.

\setlength{\unitlength}{1.8cm} 

\setlength{\unitlength}{1.8cm} 

\setlength{\unitlength}{1.8cm} 

\setlength{\unitlength}{0.8cm} 

\setlength{\unitlength}{1.8cm} 

\setlength{\unitlength}{1.8cm} 

\setlength{\unitlength}{1.8cm} 

\bigskip

\setlength{\unitlength}{1.8cm} 

\bigskip

\begin{thebibliography}{99}
\bibitem {assume}Here we assume that the scenario of the type depicted on
Fig.~\ref{fig_scenariosASEP} $b,d$ takes place.
\bibitem {other_case}With a different choice of $\rho^{A}_{R},\rho^{B}_{R}$, a
reflection graph analogous to the right graph on Fig.~\ref{fig_reflASEP} can
be obtained.
\bibitem {Naive_case}If $\mathcal{B}=I$, (\ref{mod_eigen}) becomes an
eigenvalue equation for the Jacobian ${\left(  \mathcal{D}j\right)  }$
Corresponding value of ${\frac{\delta\rho^{B}}{\delta\rho^{A}}}$ is then the
derivative along the relevant characteristic at the point $(\rho_{R}^{A}%
,\rho_{R}^{B})$.
\bibitem {strictly_speaking}Strictly speaking, the stationary density is
reached exponentially in time with characteristic time proportional to the
length of the system $N$. On the contrary, in one-species models like ASEP,
the corresponding characteristic time is of order of $1/(rate\ of\ hopping)$.
\end{thebibliography}
\end{document}